\input harvmac.tex
\input epsf.tex
\parindent=0pt
\parskip=5pt

\hyphenation{satisfying}

\def\IR{{\hbox{{\rm I}\kern-.2em\hbox{\rm R}}}}
\def\IB{{\hbox{{\rm I}\kern-.2em\hbox{\rm B}}}}
\def\IN{{\hbox{{\rm I}\kern-.2em\hbox{\rm N}}}}
\def\IC{\,\,{\hbox{{\rm I}\kern-.59em\hbox{\bf C}}}}
\def\IZ{{\hbox{{\rm Z}\kern-.4em\hbox{\rm Z}}}}
\def\IP{{\hbox{{\rm I}\kern-.2em\hbox{\rm P}}}}
\def\IH{{\hbox{{\rm I}\kern-.4em\hbox{\rm H}}}}
\def\ID{{\hbox{{\rm I}\kern-.2em\hbox{\rm D}}}}

\def\tr{{\rm tr}}

\noblackbox

\leftline{\epsfxsize1.0in\epsfbox{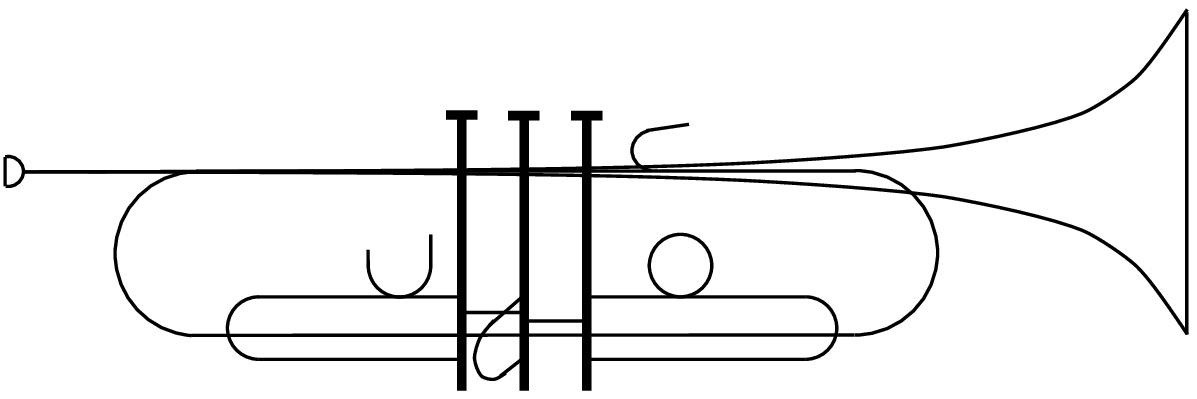}}
\vskip-0.9cm
\Title{\vbox{\baselineskip12pt
\hbox{UK/97--08}
\hbox{hep-th/9705148}}}
{On the Orientifolding of Type II NS--Fivebranes}

\centerline{\bf Clifford V. Johnson$^\dagger$}

\bigskip
\bigskip

\vbox{\baselineskip12pt\centerline{\hbox{\it Department of Physics and 
Astronomy}}
\centerline{\hbox{\it University of Kentucky}}
\centerline{\hbox{\it Lexington, KY 40506--0055 USA}}}
\footnote{}{\sl email: $^\dagger${\tt cvj@pa.uky.edu}}
\vskip1.5cm
\centerline{\bf Abstract}
\vskip0.7cm
\vbox{\narrower\baselineskip=12pt\noindent

Dualities between certain supersymmetric gauge field theories in three
and four dimensions have been studied in considerable detail recently,
by realizing them as geometric manipulations of configurations of
extended objects in type~II string theory. These extended objects
include `D--branes' and `NS--(five)branes'. In constructing the brane
configurations which realize dualities for orthogonal and symplectic
gauge groups, an `orientifold' was introduced, which results in
non--orientable string sectors.  Certain features of orientifolded
NS--branes ---such as their existence--- were assumed in the original
construction, which have not been verified directly. However, those
features fit very well together with the properties of the relevant
field theories, and subsequently yielded the known dualities. This
letter describes how orientifolded NS--branes can exist in type~II
string theory by displaying explicitly that the assumed combinations of
world--sheet and space--time symmetries do indeed leave the NS--brane
invariant and therefore can be gauged.  The resulting orientifolded
NS--brane can be described in terms of background fields, and
furthermore as an exact conformal field theory, to exactly the same
extent as the standard NS--brane.}
\vskip0.5cm

\Date{May 1997}
\baselineskip13pt

\lref\dbranes{J.~Dai, R.~G.~Leigh and J.~Polchinski,  Mod.~Phys.~Lett.
{\bf A4} (1989) 2073\semi P.~Ho\u{r}ava, Phys. Lett. {\bf B231} (1989)
251\semi R.~G.~Leigh, Mod.~Phys.~Lett. {\bf A4} (1989) 2767\semi
J.~Polchinski, Phys.~Rev.~D50 (1994) 6041, hep-th/9407031.}
\lref\orientifolds{A. Sagnotti, in {\sl `Non--Perturbative Quantum
 Field Theory'}, Eds. G. Mack {\it et. al.} (Pergammon Press, 1988), p521\semi
V. Periwal, unpublished\semi J. Govaerts, Phys. Lett. {\bf B220}
(1989) 77\semi P. Hor\u{a}va, Nucl. Phys. {\bf B327} (1989) 461.}
\lref\nsfivebrane{A. Strominger, Nucl. Phys. {\bf B343}, (1990) 167, 
{\it erratum,} {\bf 353} (1991) 565\semi
See also: C. G. Callan, J.A. Harvey and A. Strominger, 
 Nucl. Phys. {\bf 359} (1991) 611, {\it ibid.,} {\bf 367} (1991) 60.}
\lref\nscfti{S--J. Rey, in
 {\sl `Superstrings and Particle Theory: Proceedings'}, edited by
   L. Clavelli and B. Harms, (World Scientific, 1990) 291\semi
S--J. Rey, Phys. Rev. {\bf D43} (1991) 526\semi
S--J. Rey,  in 
 {\sl `The Vancouver Meeting: Particles and Fieds '91: Proceedings'}, 
edited by D. Axen, D. Bryman and  M. Comyn, (World
   Scientific, 1992) 876\semi
I. Antoniades, C. Bachas, J. Ellis and D. Nanopoulos, Phys. Lett.
 {\bf B211} (1988) 393 and Nucl. Phys. {\bf 328} (1989) 117.}
\lref\nscftii{S. Giddings and A. Strominger,  Phys. Rev. Lett. 
{\bf 67} (1992), 611.}
\lref\robdilaton{R.~C.~Myers, Phys. Lett. {\bf B199} (1987) 371.}
\lref\seiberg{N. Seiberg, Nucl. Phys. {\bf B435}, (1995) 129; hep-th/9411149.}
\lref\intriligatorthree{K. Intriligator and N. Seiberg, Phys. Lett.
 {\bf B387} (1996) 513; hep-th/9607207.}
\lref\intriligatorfour{K. Intriligator and N. Seiberg, Nucl. Phys. 
{\bf B444} (1995) 125; hep-th/9503179.}
\lref\ff{B. L. Feigin and D. B. Fuchs, Funct. Anal. Appl. {\bf 16} (1982)
 114, {\it ibid.}, {\bf 17} (1983) 241.}
\lref\wznw{S. P. Novikov, Ups. Mat. Nauk. {\bf 37} (1982) 3\semi
E. Witten, Comm. Math. Phys. {\bf 92} (1984) 455.}
\lref\rohm{R. Rohm, Phys. Rev. {\bf D32} (1984) 2849.}
\lref\wittencoset{E. Witten, Nucl. Phys. {\bf B371} (1992) 191.}
\lref\hanany{A. Hanany and E. Witten,  hep-th/9611230.}
\lref\elitzur{S. Elitzur, A. Giveon and D. Kutasov,  hep-th/9702014.}
\lref\elitzurii{S. Elitzur, A. Giveon and D. Kutasov, E. Rabinovici
 and A. Schwimmer,  hep-th/9704104.}
\lref\usthree{N. Evans, C. V. Johnson and A. D. Shapere,  hep-th/9703210.}
\lref\robin{R. W. Allen, I. Jack and D. R. T. Jones, Z. Phys. C {\bf 41} 
(1988) 323.}
\lref\witten{E. Witten, Phys. Rev. {\bf D44} (1991), 314.}
\lref\barbon{J. Barbon, hep-th/9703051.}
\lref\petr{P. Hor\u{a}va, Phys. Lett. {\bf B289} (1991) 293, hep-th/9203031.}
\lref\petrtwo{P. Hor\u{a}va, Nucl. Phys. {\bf B327} (1989) 461.}
\lref\ericjoe{E. G. Gimon and J. Polchinski, Phys. Rev. {\bf D54} (1996) 
1667, hep-th/9601038.}
\lref\joeed{J. Polchinski and E. Witten, Nucl. Phys. {\bf B460} (1996) 525,
 hep-th/9510169.}
\lref\ericme{E. G. Gimon and C. V. Johnson, Nucl. 
Phys. {\bf B477} (1996) 715, hep-th/9604129.}
\lref\robme{C. V. Johnson and R. C. Myers,  Phys. Rev. {\bf D55} (1997) 6382,
 hep-th/9610140.}
\lref\gojoe{J. Polchinski, Phys. Rev. Lett. {\bf 75} (1995) 4724, 
hep-th/9510017.}
\lref\berkooz{M. Berkooz, R. G. Leigh, J. Polchinski, J. H. Schwarz, 
N. Seiberg and E. Witten, Nucl. Phys. {\bf B475} (1996) 115,  
 hep-th/9605184.}
\lref\orbifold{L. Dixon, J. Harvey, C. Vafa and E. Witten, 
 Nucl. Phys. {\bf B261} (1985) 678;
 Nucl. Phys. {\bf B274} (1986) 285.}


{\bf 1. The NS--Brane: Background Fields}
\bigskip

The low energy limit of type II string theory describes the following 
potentials as space--time background fields:
\eqn\fieldsa{\hbox{\rm R--R}:\quad 
A^{(1)}, A^{(3)}, A^{(5)}, A^{(7)}, A^{(9)}}
for type IIA, and 
\eqn\fieldsb{\hbox{\rm R--R}:\quad 
A^{(0)}, A^{(2)}, A^{(4)}, A^{(6)}, A^{(8)}}
for type~IIB. Common to both are the potentials 
\eqn\fields{\hbox{\rm NS--NS}:\quad  B^{(2)}, B^{(6)}}
 where `R--R' refers to the sector arising from the world--sheet left
and right moving fermionic excitations which are integer moded
(Ramond) and `NS--NS' denotes those arising from the half--integer
modes (Neveu--Schwarz). The superscript denotes the rank of the
potential as an antisymmetric tensor field.  The fields denoted above
are not all independent, as ten dimensional Hodge duality relates them
by acting on their field strengths $F^{(p+2)}\leftrightarrow F^{(8-p)}$.

There are natural dynamical sources for these fields which are
$p$--dimensional extended objects (`branes') which couple electrically
to a field of rank $p{+}1$ via the $p{+}1$ dimensional world--volume
they sweep out in space--time as they propagate. In the R--R sector
the most basic\gojoe\dbranes\ electric sources are the D0--, D2--, D4--, D6--
and D8--branes, respectively for type~IIA and the D(-1)--, D1--, D3--,
D5-- and D7--branes respectively for type~IIB. By the Hodge duality
just mentioned, the D$p$--brane couples magnetically to $A^{(7-p)}$
and is thus the `dual' of the D$(6{-}p)$--brane in that sense.

Meanwhile, in the NS--NS sector the electric source for the $B^{(2)}$
field (the standard Kalb--Ramond field) is of course the fundamental
type IIA (or IIB) string itself. The electric source for $B^{(6)}$,
(and hence a magnetic source for its Hodge partner, $B^{(2)}$) is a
five dimensional extended object\nsfivebrane\ known under various
names such as `fivebrane', `NS--fivebrane', `NS--brane', `type~IIA
fivebrane', {\it etc.}, depending upon context\foot{We could
conceivably also use the term `F1--brane' and `F5--brane' as shorthand
for the fundamental string and its magnetic dual, the
NS--fivebrane. Such notation has the distinction of being of similar
form to that used for D--branes, but unfortunately is not as
completely unambiguous without clumsily appending further an `A' or a `B'
somewhere.}.

Of all of the objects described thus far, the fivebrane is the least
well understood in terms of a perturbative description. The D--branes
enjoy a very good description of their perturbative dynamics because
 their collective coordinates arise as standard open string
excitations. Meanwhile, the NS--branes are described in the closed
string sector where we can describe the collective coordinates of
extended solitonic objects much less easily. The NS--brane was
first introduced as a Yang--Mills Euclidean instanton, dressed up with
stringy fields. A direct product with six flat space--time coordinates
to make it a ten dimensional solitonic solution yielded the following
form for the background fields in type~II string theory:
\eqn\background{\eqalign{
ds^2&=-dt^2+\sum_{\alpha=1}^5dx^\alpha dx_\alpha+{\rm
e}^{2\Phi}\sum_{\mu=6}^9dx^\mu dx_\mu;\cr {\rm e}^{2\Phi}&={\rm
e}^{2\Phi_0}+{Q\over(x-x_0)^2};\cr
H_{\mu\nu\kappa}&=-\epsilon_{\mu\nu\kappa}^{\phantom{\mu\nu\kappa}\lambda}
\partial_\lambda\Phi,
}} where early Greek letters label $(t, x^1,\ldots, x^5)$, the
coordinates along the world--volume, and later ones label $(x^6, x^7,
x^8, x^9)$, the coordinates transverse to the world--volume. Also,
$x^2{=}\sum_\mu x^\mu x_\mu$.  The field $H^{(3)}$ is the rank three
field strength of $B^{(2)}$. The scalar $\Phi$ is the dilaton field,
which determines the string coupling: $g_{II}=e^\Phi$. The constant
$\Phi_0$ sets the value of the coupling at infinity. The constant
$x_0$ is the location of the centre of the fivebrane.

The number $Q$ is the $H^{(3)}$ magnetic charge of the brane, measured
by a flux integral over an asymptotic three sphere $S^3$ which
surrounds the fivebrane:
\eqn\flux{Q=-{1\over2\pi^2}\int_{S^3}H^{(3)}.} 
The charge is quantized in integer multiples of $\alpha^\prime$, the
inverse string tension, the minimal integer allowed being $1$.

One of the problems with describing this object in these terms is the
fact that at the centre of the solution, $x{=}x_0$, the dilaton field
blows up, signaling the presence of a region of strong string coupling
$g_{II}$ there. We expect that perturbation theory in $g_{II}$ departs
markedly from accuracy somewhere in the core.

\bigskip
{\bf  2. The NS--Brane: Conformal Field Theory at the Core}
\bigskip

Although the core represents a region we must treat with care, we can
approach it with some caution and learn more about it. As $x{\sim}x_0$
we can write the space--time geometry as:
\eqn\metrici{ds^2=-dt^2+\sum_{\alpha=1}^5dx^\alpha dx_\alpha+
{Q\over r^2}\left(dr^2+r^2d\Omega_3^2 \right).}  Here, we are using a
radial coordinate $r$, and $d\Omega_3^2$ is the line element on the
unit three--sphere $S^3$. We can introduce a new radial coordinate
$\sigma{=}{\rm log}_{\rm e}(r/\sqrt{Q})$ to give:
\eqn\metricii{ds^2=-dt^2+\sum_{\alpha=1}^5dx^\alpha dx_\alpha+
Q\left(d\sigma^2+d\Omega_3^2 \right),} with
\eqn\othersi{H^{(3)}=-Q\epsilon_3,\quad\quad{\rm and}\quad\Phi=-\sigma,}
where $\epsilon_3$ is the volume element on $S^3$.

So we see that after a change of variables, we can examine the theory
of the core of the NS--brane somewhat more closely.  Infinitely far
down the throat of this geometry ($\sigma{\to}-\infty$) the string
coupling, $g_{II}$, diverges, and the nature of the physics is not
clear. However, as we will recall below, the physics in the approach
to that limit is described in terms of familiar conformal field
theories, and we might hope to learn something about the theory from
this description. We will therefore (as is traditional when discussing
this limit of NS--branes) tentatively ignore our concerns about the
strong coupling region until such time as we learn more about
precisely which aspects of the physics we lose control of describing.

The theory of the non--trivial part of the core has the geometry
associated to a product of two conformal field
theories\refs{\robdilaton,\nscfti}. The $\sigma$ coordinate, with an
associated linear dilaton, is simply a Feigin--Fuchs\ff\ conformal
field theory of a free field with a background charge. The $S^3$
(angular) part, with the specified $H$--field with quantized charge
$Q{=}k\alpha^\prime$, (for $k$ integer), is precisely a conformal
field theory which can be written as an $SU(2)$
Wess--Zumino--Novikov--Witten\wznw\ (WZNW) theory at level $k$.

Alternatively, if we include the time coordinate $t$ from the
world--volume, rescale it to $\tau{=}t/\sqrt{Q}$, the $(\tau, x^6, x^7,
x^8, x^9)$ geometry of the core is the $\sigma{\to}-\infty$ limit of the
following metric:
\eqn\metriciii{ds^2=\sum_{\alpha=1}^5 dx^\alpha dx_\alpha+
Q\left(d\sigma^2-\tanh^2\!\sigma\, d\tau^2+d\Omega_3^2 \right),} with
\eqn\othersii{H=-Q\epsilon_3,\quad  \Phi={\rm log}_{\rm e}\cosh\!\sigma.}
This is simply the throat which appears at the extremal limit of a
five dimensional magnetically charged black hole. The horizon is at
$\sigma{=}0$, which is not relevant to us, as this description of the
core is only good for  larger negative $\sigma$:

This is the geometry of a product of exact conformal field theory
descriptions\nscftii. The $(t,\sigma)$ theory is an $SL(2,\IR)/U(1)$
coset at level $k$ (describing a two--dimensional black hole\witten)
and the angular parts are again the $SU(2)$ WZNW model at level
$k$. Of course, in the region where the core description is valid,
$\sigma$ large and negative, the two descriptions coincide: The
NS--fivebrane and the five dimensional black hole are indistinguishable
far down the throat.

In checking that the conformal field theories indeed give valid
superstring theory backgrounds, we should recall\robin\ that the
world--sheet fermions introduced for supersymmetry (which are
essentially free fermions for WZNW models and their
cosets\refs{\rohm,\wittencoset}), have the effect of shifting the
effective level of the WZNW models from $k$ to $k{-}2$ for the $SU(2)$
theory and to $k{+}2$ for the $SL(2,\IR)$ theory. The conformal
anomaly of the curved five dimensions of the solution is therefore
shifted from the naive
\eqn\naive{c={3k\over k+2}+{3k\over k-2}-1 +{5\over2}}  
to 
\eqn\nosonaive{c=5+{5\over2},} which is what is should be for
a total of $c{=}15$ for a complete solution, after tensoring with the
trivial conformal field theory for the $(x^1,\ldots,x^5)$ free bosons
and their superpartners.

Concentrating on the angular sector for now, let us prepare for a
later discussion by writing the standard action for the level $k$
$SU(2)$ WZNW model\wznw:
\eqn\wznwaction{S=-{k\over4\pi}\!\int_\Sigma
d^2\!z\,\, \tr[g^{-1}\partial_zg\cdot g^{-1}\partial_{\bar
z}g]-{ik\over12\pi}\!\int_{\cal
B}\!d^3\!y\,\,\epsilon^{ijk}\tr[g^{-1}\partial_ig\cdot
g^{-1}\partial_jg\cdot g^{-1}\partial_kg],} where $\Sigma$ is the two
dimensional world--sheet of the string (with topology of the sphere)
and $\cal B$ is a three dimensional ball whose boundary is $\Sigma$.
We are using complex coordinates $(z,{\bar z})$ on the world--sheet,
$g{\in}SU(2)$, and the trace is in the Lie algebra of $SU(2)$,
canonically normalized. The second term is the most natural way to
write the possible $B^{(2)}$ couplings in the theory, essentially
working in terms of the field strength $H^{(3)}$. One can always work
in terms of the potential gaining a completely two dimensional action,
but at the cost of possibly introducing `Dirac strings', and losing
some of the manifest symmetries and topological properties of the
model.

\bigskip
{\bf 3. Orientifolding}
\bigskip

In ref.\usthree, a configuration of branes in type~IIA string theory
was presented which yielded a geometrical realization of the
duality\refs{\seiberg,\intriligatorfour}\ of four--dimensional ${\cal
N}{=}1$ supersymmetric $SO(N_c)$ and $USp(N_c)$ gauge theories with
$N_f$ flavours of quarks. That configuration was an orientifold
generalization of that of ref.\elitzur, the latter yielding the
duality for $U(N_c)$. After a rotation\barbon\ and a T--duality
transformation, these configurations should imply an orientifold
generalization of the type~IIB brane configurations in
ref.\hanany. Those were designed to study dualities\intriligatorthree\
of ${\cal N}{=}4$ gauge theories in three dimensions.

The gauge sector of the four dimensional field theory is supplied by
the world--volume fluctuations of D4--branes. These branes are
suspended a finite distance between a pair of NS--branes, which are
not parallel in two of their dimensions.  As as result, the transverse
fluctuations of the D4--branes are frozen out of the problem, and the
usual hypermultiplets representing those fluctuations are not
present. The quarks are supplied by the relative fluctuations of D6--
and the D4--branes. The precise configurations of the branes are
listed in the table in the next section.

It is a natural step to go from realizing gauge group $U(N)$ on
oriented open string sectors to building gauge groups $SO(N)$ and
$USp(N)$ instead, using non--orientable string sectors. The latter may
be obtained by a projection which is known as an `orientifold'
procedure\orientifolds. 

Orientifold technology has been considerably refined in recent times,
because using it in combination with D--brane technology, many new
types of consistent string theory backgrounds may be constructed, many
of which have been indispensable in the study of string
duality\refs{\joeed,\ericjoe,\berkooz,\ericme}.

In spirit, an  orientifold is much like an orbifold. In general, it
is the gauging of a combination of a discrete spacetime symmetry 
with world--sheet parity, $\Omega$. Together, these
discrete transformations should of course form a group which we denote
as, $G_\Omega$.  $G_\Omega$ should obviously be a global
symmetry of the starting model, otherwise it would not make sense to
gauge it. Gauging $G_\Omega$ simply means to consistently project out
all states which are not invariant under the symmetry.

Orientifolding has been largely carried out in the context of either
toroidal compactifications, or orbifolds thereof, and so not much is
known about orientifolding closed superstring backgrounds with a
non--trivial distribution of curvature.  In trying to extend the work
of ref.\elitzur\ to incorporate $SO(N)$ or $USp(N)$ gauge groups in
the open string sector, the authors of ref.\usthree\ were forced to
consider the action of the orientifold group on the NS--brane.
 
An immediate concern about orientifolding type~II theory is the
question of the very existence of a fivebrane in the resulting
theory. In the most simple orientifold model, type~I string theory,
there is no NS--brane. This is because there is no $B^{(6)}$ potential
for it to couple to. This is easy to see if one considers the basic
$\sigma$--model coupling of fundamental type II~strings to $B^{(2)}$:
\eqn\coupling{\int\!d^2\!z\, B_{\mu\nu}\left(\partial_z
x^\mu\partial_{\bar z}x^\nu-\partial_{\bar
z}x^\mu\partial_zx^\nu\right).}  The action of world--sheet parity is
$\Omega: z{\leftrightarrow}{\bar z}$, and therefore $B^{(2)}$ must go
to $-B^{(2)}$ under~$\Omega$ in order for $\Omega$ to be a
symmetry. In type~IIB string theory, where the left and right
supersymmetries are of the same chirality, the group
$G_\Omega{=}\{1,\Omega\}$ is a symmetry of the theory and may be
gauged. In the process, $B^{(2)}$ and its Hodge partner $B^{(6)}$ are
projected out of the theory, as they are odd. There are no NS--branes
in the resulting type~I theory.

The analogous minimal case for type~IIA string theory is to consider
the combination of $\Omega$ with a reflection in one of the
space--time coordinates (say $x^9$). The reflection,
$R_9:~x^9{\to}{-}x^9$, mirrors the two space--time supersymmetries into
each other, and $G_\Omega{=}\{1,\Omega R_9\}$ is a global symmetry which
can therefore be gauged, resulting in the type~I$^\prime$
theory\foot{Which should perhaps have  been called the `type IA'
theory, and its $T_9$--dual cousin the `type IB' theory.}, which is of
course $T_9$--dual to the type~I theory.

The type~I$^\prime$ theory has 16 D8--branes, giving $SO(32)$ gauge
symmetry, and an `O8--plane', located at a point in the $x^9$
direction, about which the type~IIA theory is reflected. The D8--branes
are $T_9$--dual to the D9--branes defining type~I theory, and the
O8--plane is the space--time manifestation of combining T--duality
with orientifolding: $\Omega{\leftrightarrow}\Omega R_9$: It is the
fixed point set of the $R_9$ reflection.

By using T--duality on such flat backgrounds as above, we can
deduce therefore that the world--sheet fermions (and the corresponding
space--time supersymmetry they generate after the GSO projection)
constrain us to consider consistent gauging of only $\Omega$ times an
{\sl even} number of space--time reflections for type~IIB; and times
only an {\sl odd} number of space--time reflections for type~IIA. This
gives O$n$--planes in each theory with $n$ odd for type~IIB and even
for type~IIA.  By extension, we can expect that this restriction will
be applicable to any non--trivial backgrounds which are asymptotically
flat, which includes the NS--branes under consideration.

\bigskip
{\bf 4. Orientifolding NS--Branes}
\bigskip

To proceed, we must learn from the example of the type~IIA orientifold
and find an orientifold group that is complicated enough to produce
symmetries of the action which allow the couplings to the NS--NS
potential to survive at least in some sectors of the theory, allowing
the possibility for an NS--brane to be defined.

In ref.\usthree, it was observed that the only type of orientifolds of
the configuration of ref.\elitzur\ which would preserve the existing
$D{=}4$, ${\cal N}{=}1$ supersymmetry were of the type $\{1,\Omega
R_{45789}\}$ and $\{1,\Omega R_{456}\}$. (Here $R_{\mu\nu}{\equiv}R_\mu
R_\nu$, {\it etc.}) This introduces O4--planes and O6--planes
respectively.

A table showing the  configuration of branes and {\sl possible}
 orientifolds is given below: 
\bigskip
\vbox{
$$\vbox{\offinterlineskip
\hrule height 1.1pt
\halign{&\vrule width 1.1pt#
&\strut\quad#\hfil\quad&
\vrule width 1.1pt#
&\strut\quad#\hfil\quad&
\vrule#
&\strut\quad#\hfil\quad&
\vrule#
&\strut\quad#\hfil\quad&
\vrule#
&\strut\quad#\hfil\quad&
\vrule#
&\strut\quad#\hfil\quad&
\vrule#
&\strut\quad#\hfil\quad&
\vrule#
&\strut\quad#\hfil\quad&
\vrule#
&\strut\quad#\hfil\quad&
\vrule#
&\strut\quad#\hfil\quad&
\vrule#
&\strut\quad#\hfil\quad&
\vrule width 1.1pt#\cr
height3pt
&\omit&
&\omit&
&\omit&
&\omit&
&\omit&
&\omit&
&\omit&
&\omit&
&\omit&
&\omit&
&\omit&
\cr
&\hfil type&
&\hfil $t$&
&\hfil $x^1$&
&\hfil $x^2$&
&\hfil $x^3$&
&\hfil $x^4$&
&\hfil $x^5$&
&\hfil $x^6$&
&\hfil $x^7$&
&\hfil $x^8$&
&\hfil $x^9$&
\cr
height3pt
&\omit&
&\omit&
&\omit&
&\omit&
&\omit&
&\omit&
&\omit&
&\omit&
&\omit&
&\omit&
&\omit&
\cr
\noalign{\hrule height 1.1pt}
height3pt
&\omit&
&\omit&
&\omit&
&\omit&
&\omit&
&\omit&
&\omit&
&\omit&
&\omit&
&\omit&
&\omit&
\cr
&\hfil NS&
&\hfil --- &
&\hfil --- &
&\hfil --- &
&\hfil --- &
&\hfil --- &
&\hfil --- &
&\hfil $\bullet$ &
&\hfil $\bullet$ &
&\hfil $\bullet$ &
&\hfil $\bullet$ &
\cr
height3pt
&\omit&
&\omit&
&\omit&
&\omit&
&\omit&
&\omit&
&\omit&
&\omit&
&\omit&
&\omit&
&\omit&
\cr
\noalign{\hrule}
height3pt
&\omit&
&\omit&
&\omit&
&\omit&
&\omit&
&\omit&
&\omit&
&\omit&
&\omit&
&\omit&
&\omit&
\cr
&\hfil NS$^\prime$&
&\hfil --- &
&\hfil --- &
&\hfil --- &
&\hfil --- &
&\hfil $\bullet$ &
&\hfil $\bullet$ &
&\hfil $\bullet$ &
&\hfil $\bullet$ &
&\hfil --- &
&\hfil --- &
\cr
height3pt
&\omit&
&\omit&
&\omit&
&\omit&
&\omit&
&\omit&
&\omit&
&\omit&
&\omit&
&\omit&
&\omit&
\cr
\noalign{\hrule}
height3pt
&\omit&
&\omit&
&\omit&
&\omit&
&\omit&
&\omit&
&\omit&
&\omit&
&\omit&
&\omit&
&\omit&
\cr
&\hfil O4&
&\hfil --- &
&\hfil --- &
&\hfil --- &
&\hfil --- &
&\hfil $\bullet$ &
&\hfil $\bullet$ &
&\hfil --- &
&\hfil $\bullet$ &
&\hfil $\bullet$ &
&\hfil $\bullet$ &
\cr
height3pt
&\omit&
&\omit&
&\omit&
&\omit&
&\omit&
&\omit&
&\omit&
&\omit&
&\omit&
&\omit&
&\omit&
\cr
\noalign{\hrule}
height3pt
&\omit&
&\omit&
&\omit&
&\omit&
&\omit&
&\omit&
&\omit&
&\omit&
&\omit&
&\omit&
&\omit&
\cr
&\hfil O6&
&\hfil --- &
&\hfil --- &
&\hfil --- &
&\hfil --- &
&\hfil $\bullet$ &
&\hfil $\bullet$ &
&\hfil $\bullet$ &
&\hfil --- &
&\hfil --- &
&\hfil --- &
\cr
height3pt
&\omit&
&\omit&
&\omit&
&\omit&
&\omit&
&\omit&
&\omit&
&\omit&
&\omit&
&\omit&
&\omit&
\cr
\noalign{\hrule}
height3pt
&\omit&
&\omit&
&\omit&
&\omit&
&\omit&
&\omit&
&\omit&
&\omit&
&\omit&
&\omit&
&\omit&
\cr
&\hfil D4&
&\hfil --- &
&\hfil --- &
&\hfil --- &
&\hfil --- &
&\hfil $\bullet$ &
&\hfil $\bullet$ &
&\hfil [---] &
&\hfil $\bullet$ &
&\hfil $\bullet$ &
&\hfil $\bullet$ &
\cr
height3pt
&\omit&
&\omit&
&\omit&
&\omit&
&\omit&
&\omit&
&\omit&
&\omit&
&\omit&
&\omit&
&\omit&
\cr
\noalign{\hrule}
height3pt
&\omit&
&\omit&
&\omit&
&\omit&
&\omit&
&\omit&
&\omit&
&\omit&
&\omit&
&\omit&
&\omit&
\cr
&\hfil D6&
&\hfil --- &
&\hfil --- &
&\hfil --- &
&\hfil --- &
&\hfil $\bullet$ &
&\hfil $\bullet$ &
&\hfil $\bullet$ &
&\hfil --- &
&\hfil --- &
&\hfil --- &
\cr
height3pt
&\omit&
&\omit&
&\omit&
&\omit&
&\omit&
&\omit&
&\omit&
&\omit&
&\omit&
&\omit&
&\omit&
\cr
}\hrule height 1.1pt
}
$$
}

\bigskip

In the table, a dash `---' represents a direction {\sl along} an
extended object's world--volume while a dot~`$\bullet$' is
transverse. For the special case of the D4--branes' $x^6$ direction,
where a world--volume is a finite interval, we use the symbol `{\rm
[---]}'.  (A `$\bullet$' and a `---' in the same column indicates that
one object is living inside the world--volume of the other in that
direction, and so they can't avoid one another. Two `$\bullet$'s in
the same column tell us that the objects are point--like, and need not
coincide in that direction, except for the specific case where they
share identical values of that coordinate.)

Let us consider the first NS--brane in the table, point--like in the
$(x^6, x^7, x^8, x^9)$ directions, like our NS--brane in the previous
sections. All of our comments will apply equally well to the second,
differently oriented one, which is denoted NS$^\prime$ in the table.

In the case where we have an O4--plane present, the spacetime
reflection is $R_{45789}$. The action of reflections in the $(x^4,
x^5)$--plane, where the NS--brane has no structure, will produce no
physics of interest for us, so it will suffice to study
$R_{789}$. Notice that the reflections are only in three of the four
directions in which the brane has structure.

In the case where we have an O6--plane present, the non--trivial
reflection acting on the NS--brane is $R_6$. Now, this is a
reflection on just one of the four directions in which the brane has
structure.

In order to proceed, it is prudent to change coordinates from the
Cartesian ones we have been using for transverse directions to radial
and angular coordinates $(r,\phi,\psi,\theta)$. For this, we can use
Euler coordinates:
\eqn\euler{\eqalign{
x^6&=r\cos\left({\phi+\psi\over2}\right)\cos\!{\theta\over2}\cr
x^7&=r\sin\left({\phi+\psi\over2}\right)\cos\!{\theta\over2}\cr
x^8&=r\cos\left({\phi-\psi\over2}\right)\sin\!{\theta\over2}\cr
x^9&=r\sin\left({\phi-\psi\over2}\right)\sin\!{\theta\over2},
}}
where $0\leq\theta<\pi$, $0\leq\phi<2\pi$ and $0\leq\psi<4\pi$.

In these coordinates, the metric on the unit three--sphere is:
\eqn\metricsthree{d\Omega_3^2=d\theta^2+d\psi^2+d\phi^2
+2\cos\!\theta d\phi d\psi,} while the volume element is:
\eqn\volume{\epsilon_3={1\over4\pi}\sin\!\theta\, d\theta d\phi d\psi.}

Therefore, in the core limit, the $\sigma$--model coupling of 
$B^{(2)}$ must be:
\eqn\coupling{B_{\phi\psi}
\left(\partial_z\phi\partial_{\bar z}\psi-
\partial_{\bar z}\phi\partial_z\psi\right),}
where
\eqn\bfield{B_{\phi\psi}={Q\over4\pi}(\pm1-\cos\!\theta),}
where the $\pm$ choice refers to the North and South poles,
$\theta=0$ and $\pi$, respectively.

Now let us study the action of $R_{789}$ and $R_6$ in
these coordinates.  Upon examination of the coordinate transformations
\euler\ above, we see that:
\eqn\flips{\eqalign{R_{789}:&\,\, 
\biggl\{\matrix{\psi\to-\pi-\phi\cr\phi\to\phantom{-}\pi-\psi}
 \biggr.\cr
R_6:&\,\, 
\biggl\{\matrix{\psi\to\pi-\phi\cr\phi\to\pi-\psi}
 \biggr.
}}
leaving all other coordinates changed in each case.

Looking at the coupling of the NS--brane's $B^{(2)}$--field, we see
that in each case, combining the reflection with parity
$\Omega:~z\leftrightarrow{\bar z}$ leaves the combination
$(\partial_z\phi\partial_{\bar z}\psi-\partial_{\bar
z}\phi\partial_z\psi)$ invariant. Therefore, we need to have $B^{(2)}$
be {\sl even} under $\Omega R_{789}$ or $\Omega R_6$ which it is,
because of its antisymmetry under exchange of indices and by virtue of
it being odd under $\Omega$.

Notice that we could have also considered $\Omega R_{6789}$ and
$\Omega R_{89}$, thus acting only on an even number of directions
transverse to the fivebrane. In these cases we have:
\eqn\flips{\eqalign{R_{89}:&\,\, 
\biggl\{\matrix{\psi\to\psi-\pi\cr\phi\to\phi+\pi}
 \biggr.\cr
R_{6789}:&\,\, 
\biggl\{\matrix{\psi\to\psi-2\pi\cr\!\!\phi\to\phi\phantom{+2\pi}}
 \biggr. }} again leaving all other coordinates changed in each case.
So we see that $R_{89}$ and $R_{6789}$ are manifestly symmetries {\sl
before} acting with $\Omega$ and therefore will result in $B^{(2)}$
being odd under $\Omega R_{89}$ and $\Omega R_{6789}$, and hence it
will be projected out in constructing the gauged theory.

Notice that at this level it seems to be consistent to have an
orientifold with {\sl both} O6-- and O4--planes present, in the way
they are aligned in the table. (Such configurations were considered
recently in ref.\elitzurii, following ref.\usthree.) The candidate
orientifold group $G_\Omega$ would have elements $\Omega R_{45789}$
and $\Omega R_{456}$. By closure of the group algebra (and necessarily
the operator product expansion) this leads to the presence of the pure
orbifold symmetry $R_{6789}$, which is just the overall reflection
symmetry $\psi{\to}\psi{-}2\pi$ of the model.

Although we have discussed the detailed form of $B^{(2)}$ only in the
core limit, our conclusions about the global symmetries extend to the
complete NS--brane, as the asymptotic region is smoothly connected to
the throat region of the core, developing a less trivial
dependence on the radial coordinate in the fields (see \background) as
this happens.  At any radius, the $SO(4)$ rotational symmetry
guarantees that we can always do the decomposition and analysis of
this section, and so we conclude that the couplings of the {\sl
complete NS--brane} admit the operations $\Omega R_{6}$ and $\Omega
R_{789}$ as global symmetries.

\bigskip
{\bf 5. The Orientifolded NS--Brane: The Theory at the Core}
\bigskip

In the  WZNW model, the interesting reflection symmetries have a very
natural interpretation in terms of the group element $g$. In Euler
coordinates, it is written:
\eqn\g{g={\rm e}^{i\phi\sigma_3/2}
{\rm e}^{i\theta\sigma_2/2}{\rm e}^{i\psi\sigma_3/2}
=\pmatrix{{\rm e}^{i{\phi+\psi\over2}}
\cos\!{\theta\over2}&{\rm e}^{i{\phi-\psi\over2}}\sin\!{\theta\over2}
 \cr -{\rm e}^{-i{\phi-\psi\over2}}\sin\!{\theta\over2}& {\rm
e}^{-i{\phi+\psi\over2}}\cos\!{\theta\over2}}} (where the $\sigma_i$
 are the Pauli matrices), from which we see:
\eqn\gaction{\eqalign{R_6:\,\,& g\to
 -g^{-1}\cr R_{789}:\,\,&
g\to\phantom{-} g^{-1}.}}  Upon combining these with $\Omega:
 z{\leftrightarrow}{\bar
z}$, these are  manifestly symmetries of the WZNW action~\wznwaction.

It would be very interesting to carry out the complete study of this
gauged model, defining a new theory with non--orientable sectors.
Notice, for example that the conserved left-- and right--moving
Lie algebra valued currents,
\eqn\currents{J(z)=kg^{-1}\partial_zg 
\quad\quad{\rm and}\quad {\bar J}({\bar z})
=k\partial_{\bar z}gg^{-1},} in terms of which much of the physics of
the conformal field theory is determined, are mapped into each other
under the orientifold groups\foot{It should be noted here that an
early example of a study of a non--trivial orientifold was carried out
in ref.\petr, in the context of gauging spacetime symmetries of the
two dimensional geometry of the (bosonic) stringy black hole
background defined by the $SL(2,\IR)/U(1)$ coset.}.

\bigskip
\bigskip
{\bf 6. Closing Remarks}
\bigskip

In closing, we have demonstrated that the assumption made in
ref.\usthree\ was correct. It is indeed possible to define a
consistent orientifold projection on the theory of an NS--brane by
combining world--sheet parity, $\Omega$, with certain spacetime
reflections: We can introduce orientifold planes where  one or
three of the four transverse NS--brane directions are also transverse
to the orientifold plane. As there is no non--trivial structure in the
directions along the NS--brane's world--volume, it is clear by
T--duality that we can have orientifold planes with $n$ extended
directions from $n=1,\ldots,8$ intersecting the NS--brane
consistently, providing that only one or three transverse coordinates
are acted upon. Taking into account our earlier restriction (due to
spacetime supersymmetry) on the number of reflections allowable, $n$
has to be even in type~IIA and odd in type~IIB, just like the brane
dimensions.

This demonstration serves to identify the global orientifold
symmetries, denoted $G_\Omega$, which should be gauged to find the
projected theory. It would be an interesting task to study the full
projected conformal field theory. This is a very tractable problem,
for the WZNW model organizes the conformal field theory very well. It
will be particularly interesting to study such orientifolded
NS--branes in a compact situation where associated fluxes cannot leak
off to infinity. There, additional consistency checks will arise from
computing the Klein bottle diagram which now appears in the theory at
one loop. Any tadpole divergences associated with this diagram might
conceivably be canceled with cylinder and M\"obius--strip diagrams
arising from introducing D--branes to obtain consistent
backgrounds. In this way, a whole new class of
interesting non--trivial orientifold $+$ D--brane $+$ NS--brane
backgrounds can be constructed which will certainly be interesting for
studies of the properties of both string and field theory.

\bigskip
\medskip

\noindent
{\bf Acknowledgments:}

\noindent
CVJ was supported by family, friends and music. Thanks to Robert~C.
Myers for comments on the manuscript.

\bigskip
\bigskip

\centerline{\epsfxsize1.0in\epsfbox{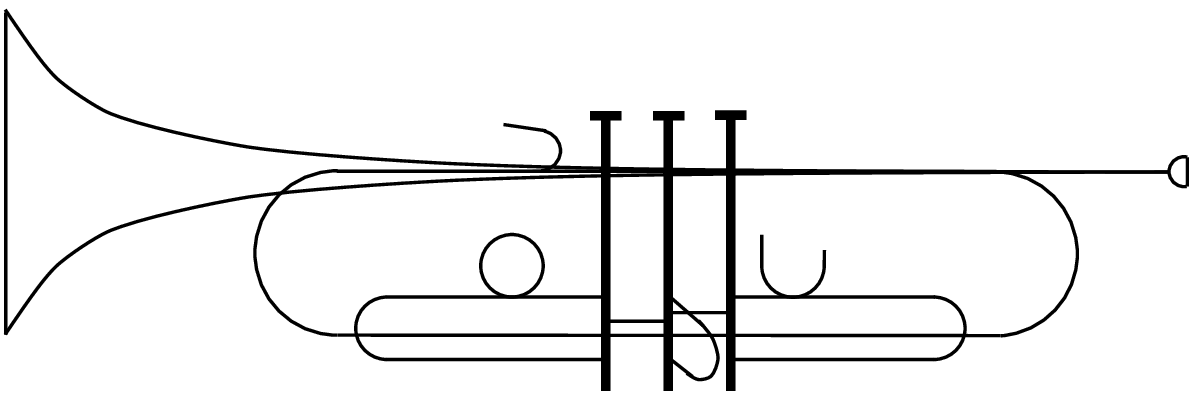}}

\listrefs

\bye